# Computational snapshot angular-spectral lensless imaging

P. Wang and R. Menon[a)]

*Department of Electrical and Computer Engineering, University of Utah, Salt Lake City, UT 84112, U.S.A.*

**Abstract**: By placing a diffractive element in front of an image sensor, we are able to multiplex the spectral and angular information of a scene onto the image sensor. Reconstruction of the angular-spectral distribution is attained by first calibrating the angular-spectral response of the system and then, applying optimization-based matrix inversion. In our proof-of-concept demonstration, we imaged the 1D angle and the spectrum with resolutions of $0.15^o$ and 6nm, respectively. The information is reconstructed from a single frame, thereby enabling snapshot functionality for video-rate imaging.

1. **Introduction**

Conventional cameras image intensity distribution of an object. However, the complete description of the lightfield also contains the direction information of light rays [1]. Therefore, lightfield is essentially represented by the 5-dimensional intensity data **I**($x$, $y$, $z$, $θ$, $φ$), in which ($x$, $y$, $z$) represents the 3-dimensional spatial coordinates of a point and ($θ$, $φ$) in spherical coordinates represents the direction of a light ray emanating from that point. Previously, either large camera array [2,3] or microlens array [4,5] were utilized to construct angle-sensitive cameras. They were composed of multiple optical lenses or sensors, thus they are bulky and expensive. With its applications in volumetric imaging and numerical re-focusing [4,6,7], there is a need to simplify the imaging system by reducing its form factor. An angle-sensitive sensor is able to record light field, but it is still a lens-based system and it requires a special sensor [8]. Illumination using an array of LEDs can also create the angular-spatial data [9]. Angle-sensitive cameras are

---

[a)] Electronic mail: rmenon@eng.utah.edu

essentially enabled by both the construction of optical systems and the development of computational algorithms.

Since a semiconductor sensor converts photons to electrons without discriminating wavelength of light, color-filter-array (CFA), such as Bayer filter, is widely adopted for color imaging. Nevertheless, only three basic colors (red, green and blue) can be detected. In order to sample object spectra **I**(*x*, *y*, *λ*) in finer resolution, multi-spectral and hyper-spectral imagers were invented. One of these, the pushbroom method has limited throughput due to 1D imaging and mechanical scanning [10]. Tunable filter is costly and requires multiple exposures to record one hyper-spectral cube **I**(*x*, *y*, *λ*) [10]. Recent advances in nanofabrication enabled tiled-filter-array sensors, which unfortunately compromises spatial resolution and suffers from degraded light sensitivity due to absorption [11,12]. These methods, which can be classified as division-of-focal-plane suffers from reduced field-of-view and photon throughput. They also experience parallax, which complicates post-processing of the images. Coded aperture is a promising alternative, which however is subject to reduced image quality and relatively low light throughput since the code typically has less than 50% transmission efficiency [13].

In this paper, a snapshot angular-spectral lensless imaging system is described and experimentally demonstrated. Its sensitivity to angle and wavelength is enabled by transmission through a diffractive element placed in front of a conventional image sensor. The diffractive element is patterned by single-step grayscale lithography. The diffraction pattern of an incident angle and a wavelength band serves as a unique fingerprint for that angle and wavelength band. The 2D monochrome image received at the sensor plane is essentially a linear superposition of a set of diffraction patterns of all the angles and wavelengths that constitute the original object. A computational algorithm is successfully implemented to recover the multiplexed multi-dimensional information from a single-shot sensor image. One primary advantage is its small footprint, since no optical lens is needed for imaging [14,15]. This is critical for applications where cost and space may be major concerns. Another advantage comes from the fact that both angular and spectral distributions of light rays are extracted from a single snapshot without any scanning. This is important for applications that probe fast dynamics and motion blur is undesired [10]. It is important to note that our current implementation is not a lightfield camera and as such, depth information cannot be obtained. The complete hyperspectral lightfield data requires calibration in all 6 dimensions, (*x*, *y*, *z*, *θ*, *φ*, *λ*). In our current implementation, we performed calibration and reconstruction only in 2 dimensions for simplicity, $(\theta, \lambda)$. However, the working principle described below can be extended to the full 6 dimensional hyper-spectral lightfield.

## 2. Working principle

The basic principle is illustrated in Fig. 1, where a sensor array is placed at a distance $d$ from the diffractive optic. Different wavelengths generate different diffraction patterns (blue, green and red arrows and lines). This approach was previously utilized in computational spectroscopy with high bandwidth-to-resolution ratio [16,17] and a computational multi-spectral camera [18]. In addition to reconstructing spectrum, in this work, we further explore the sensitivity of the diffractive optic to the angle of incidence. As illustrated in Fig. 1, the diffraction pattern on the sensor is sensitive to not only the wavelength, but also the angle of incidence. Furthermore, any scene may be decomposed into a group of 3D spatial points $(x, y, z)$ and wavelengths $(\lambda)$. In this decomposition, each spatial point of the object emits a spherical wave. Since spherical waves contain a rich set of spatial frequencies $(\theta_x, \theta_y)$, we can alternatively represent the object in the spatial-frequency domain, or in another words, the angular domain, i.e., $\mathbf{I}(\theta_x, \theta_y, \lambda)$. Any scene can be represented as a linear combination of basis functions from the spherical waves centered on an object point and of a given wavelength. The diffractive optic creates a unique response to each of these basis functions on the image sensor because of its sensitivity to angle and wavelength. If these responses are first recorded, then it becomes possible to perform numerical inversion to ascertain the spectral-angular response of any general object or scene as described below.

The angular-spectral distribution $\mathbf{I}(\theta_x, \theta_y, \lambda)$ of light rays from the object can be reconstructed from one single monochrome sensor image with the help of an inversion algorithm. The origins of point sources that constitute the object can be extracted based on simple geometric relations, which is shown later. For simplicity, here we only consider 1D imaging (Fig. 1), which is simplified to $\mathbf{I}(\theta_x, \lambda)$ or $\mathbf{I}(\theta, \lambda)$. In other words, we utilize a 1D diffractive element to demonstrate a 1D angular-spectral imager. A simulation of imaging a 2D object is carried out to illustrate the potential of 2D angular-spectral imaging.

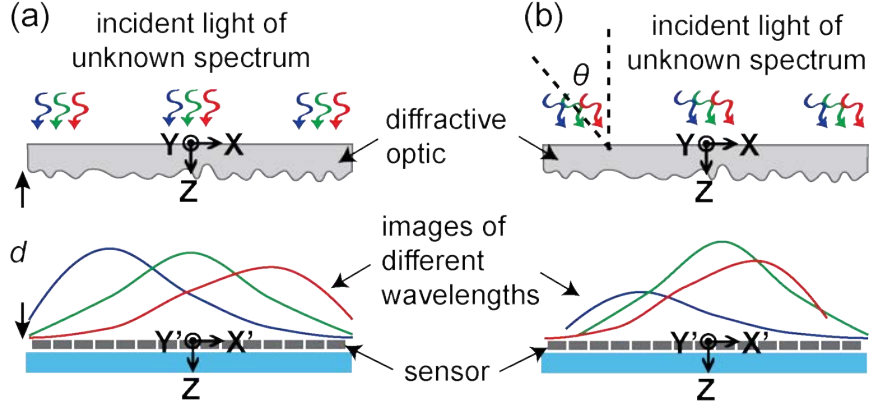

Fig. 1. Working principle of the computational angular-spectral imaging system. The diffraction pattern on the image sensor after transmitting the diffractive optic is sensitive to both the angle of incidence ($\theta$) and wavelength ($\lambda$) (a) Normal incidence; (b) Oblique incidence.

The basic idea is to reconstruct the angular-spectral distribution $\mathbf{I}(\theta, \lambda)$ of the object from a single 1D monochrome image $\mathbf{S}(x')$. Specifically, a 2D image is first captured and then averaged along the degenerate direction (Y') to get the 1D image. The unknown angular-spectral image, $\mathbf{I}$ and the measured sensor data, $\mathbf{S}$ are related via a system matrix, $\mathbf{A}$, which contains diffraction patterns of many angular frequencies ($\theta$) and wavelengths ($\lambda$). These diffraction patterns are also 1D monochrome images and can be considered as angular-spectral point-spread-functions (AS-PSFs). The forward model is simply expressed as $\mathbf{S} = (\mathbf{AQ})\mathbf{I}$, where $\mathbf{Q}$ is the diagonal matrix of quantum efficiency versus wavelength. Direct matrix inversion $\mathbf{I} = (\mathbf{AQ})^{-1}\mathbf{S}$ is usually impractical, since $\mathbf{A}$ is typically ill-conditioned. Either iterative or non-iterative methods may be applied. In this work, a modified version of direct-binary-search (DBS) algorithm was implemented to solve this inverse problem [16,18,19].

## 3. Diffractive optic

The diffractive optic used here is comprised of multiple pixels of uniform width (3μm). It has a staircase profile, where each pixel is assigned a specific height. For proof-of-principle only 1D design is employed in this work [18]. As depicted in Fig. 1, the optic has topography variation along X direction, while uniform along Y direction. Therefore, the diffraction pattern on the sensor is sensitive to changes of incident angle in the XZ plane. A commercially available grayscale lithography machine was employed to fabricate this diffractive element [20]. To start with, a thin layer of positive photoresist (Shipley 1813) was uniformly coated on a double-side-polished glass substrate using a spinner. It is then soft baked in an oven at 110$^o$C for 30min. The thin film has a thickness of 1.8μm. Next, the photoresist is exposed in a laser-pattern generator (Heidelberg microPG101), which works in the direct-laser-writing mode and modulates the exposure dose at different pixels in grayscale. Eventually, the photoresist is developed in 352 developer for 1 min. A photograph of the patterned diffractive optic on a glass substrate is shown in Fig. 2(a). The final device covers a total area of 18×18mm$^2$. It has 3 repeated designs and each one is 6mm in width. A closer look at the microstructures by optical microscope clearly shows its 1D pixels (or grooves). Figure 2(c) is the topography measured using atomic force microscopy. It scanned a 30μm by 12μm area, covering 10 pixels. The maximum depth is 1.2μm. Similar diffractive optics was formerly used in photovoltaics [20] and microscopy [21].

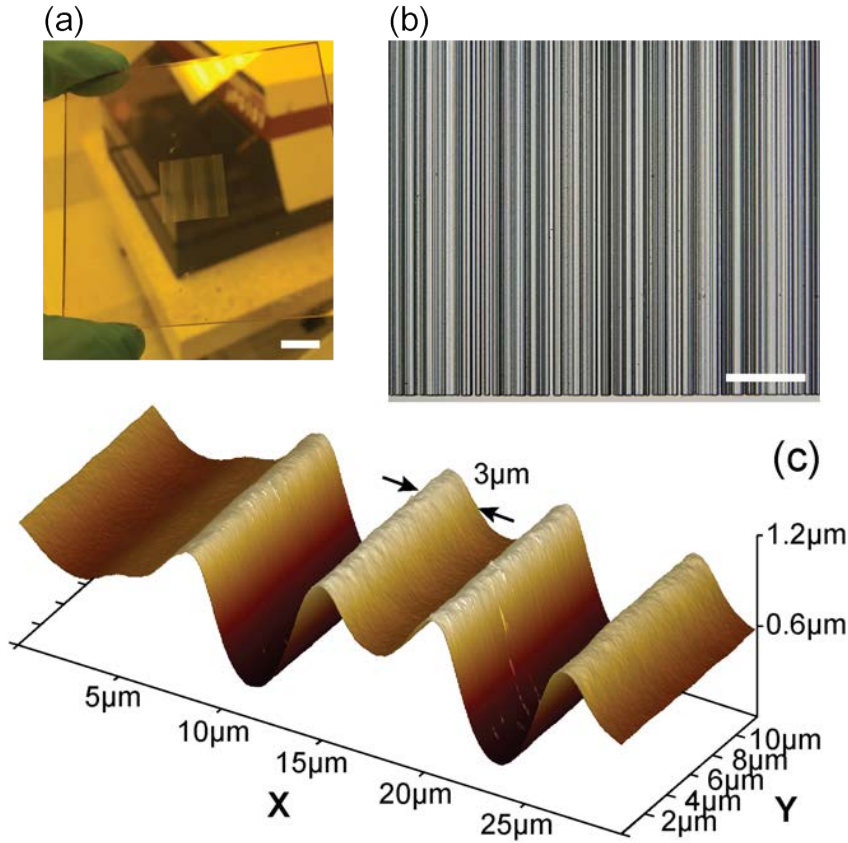

Fig. 2. (a) Photograph of the fabricated diffractive optic on a 3" glass substrate (scale bar: 1cm). (b) Optical microscope image (scale bar: 100um). (c) AFM measurement of a small segment.

Since both the photoresist and the substrate are highly transparent in the visible band, the only major losses are due to Fresnel reflections at the interface, leading to high photon throughput. Fresnel reflection (~4%) can be minimized by an anti-reflection coating (ARC) [22]. In our system, only one surface needs to be coated, compared to other lens-based systems, which incorporate multiple lenses. Many alternative multi-spectral imagers suffer from low photon throughput. For instance, the tiled-filter-array only uses a small fraction of photons due to filter absorption [11,12] and coded aperture blocks at least half the incident light due to the amplitude mask [13].

The diffractive optic used in our experiments reported here was originally designed for multi-color fluorescence microscopy and its design has been reported previously [21]. Its topography was optimized using the direct-binary-search (DBS) algorithm. All its pixels were perturbed in each iteration to improve the figure-of-merit (FOM).

## 4. Calibration

To solve for the aforementioned inverse problem, it is imperative to fully characterize the system matrix, **A**, which contains the diffraction patterns (or AS-PSFs) of all wavelengths and angles. Figure 3 shows the schematic of the calibration setup. Here, broadband light from a super-continuum source (SuperK COMPACT, NKT Photonics) is first conditioned by two lenses, separated by a distance of the sum of their focal lengths ($d_0 = f_1 + f_2$). When both lenses are aligned on the optical axis, the diffractive optic sees light of normal incidence. If the first lens mechanically scans along the X direction, different oblique angles can be created on the exit side of the second lens. The angle $\theta$ is defined by the movement of the first lens: $\theta \approx \tan(\theta) = \Delta x / f_2$. $\Delta x$ represents the displacement of lens 1 with respect to the optical axis. To characterize the diffraction patterns at different wavelengths, a single-mode-fiber-tip is mounted on a second scanning stage (along the X' direction), at gap, $d$=90mm from the diffractive optic. It feeds the light signal at different locations to a spectrometer (Ocean Optics Jaz). Thereby, this custom-built system is able to record the angular-spectral PSFs. Note that the angular response of the fiber is assumed to remain flat over the range of angles that we used in our experiments. An alternative way of calibration is to filter the broadband light source using bandpass filters and take sensor images. However, we did not pursue this approach due to two reasons. First, it is expensive and cumbersme to have bandpass filters with very narrow bandwidths (5nm) and consecutive central wavelengths over the entire visible spectrum. Second, our spectrometer has a higher dynamic range than our sensor.

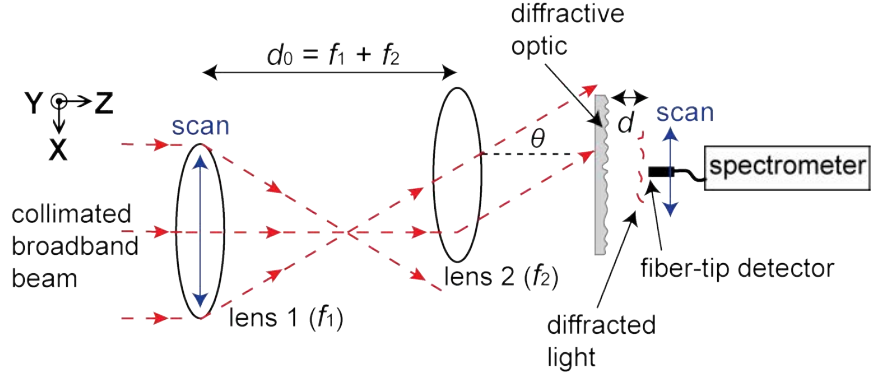

Fig. 3. Schematic of the angular-spectral PSF calibration setup. Different angles of incidence are created by scanning the first lens.

Figure 4 summarizes three examples of the calibration results for three example angles - 2.56°, 0° and +3.84°. The horizontal direction X' stands for the spatial domain at the sensor plane and the vertical direction corresponds to the spectral domain. The bright oblique lines represent the intensity distribution as a function of wavelength and incident angle on the image sensor after transmission through the diffractive optic. As expected, different angles of incidence, $\theta$ (Figs. 4(a), (b) and (c)) give different spatial-spectral intensity maps, or diffraction patterns. Similarly, different wavelengths also have different intensity distributions along X'. On the contrary, the faint vertical lines are almost independent of wavelength. These are essentially background noise due to fabrication errors and do not contribute to image reconstruction. The spectrum covers the entire visible band. An IR cut filter is used to eliminate infrared from the super-continuum source. Note that the spectrometer has a spectral resolution of 0.4nm. However, a spectral sampling rate of 5nm in reconstruction is assumed by interpolating the raw data. The scan along X' direction has a length of 6mm with 10μm step, dictated by the micrometer stage. As mentioned above, the step of angular scan is determined by the displacement of the first lens. Here $\Delta x$ = 500μm was used. A lens focal length of $f_2$ = 90mm gives an angular scan step of 0.32°. A total scan from -7.5mm to 7.5mm corresponds to angles from -4.8° to 4.8°. Thus there are 31 angles in total. Note that the super-continuum beam is well collimated and expanded to more than 2" in diameter so that it covers the entire range of scan of lens 1. The beam also has properly homogenized intensity profile to eliminate any effects from non-uniform incident light.

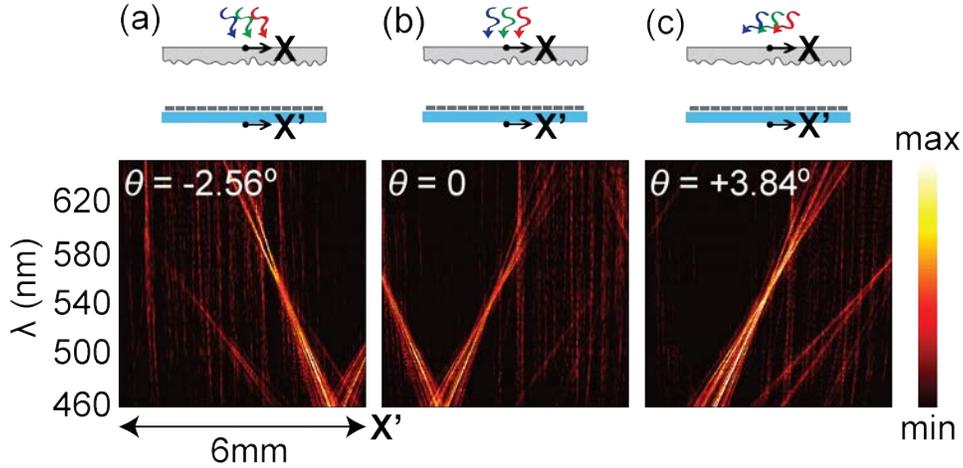

Fig. 4. Top row: Schematic of different angles of incidence. Bottom row: Spatial-spectral diffraction patterns measured at three angles: (a) -2.56°; (b) normal incidence; (c) +3.84°.

## 5. Results and discussions

5.1 Experimental demonstration

A simple scene was imaged for demonstration. It contains two light-emitting-diodes (LEDs) of two colors, placed at two distinct locations. These can be approximated as point sources. A green LED with central wavelength around 518nm is placed at 100mm from the diffractive optic. A red LED with central wavelength around 625nm is 150mm and 2mm away from the green LED in the longitudinal and transverse directions, respectively. This configuration is sketched in Fig. 5(a). Figure 5(b) gives the raw monochrome image captured by our CCD sensor (Andor Clara). Note that this sensor was used only due to its availability and our approach is agnostic to the sensor. The angular sensitivity of the sensor is assumed constant, since the range of angles here is much smaller than its cut-off angle. The plot of the 1D image is overlaid on the raw image. As discussed earlier, this intensity variation represents the linear combination of the diffraction patterns of all the wavelengths and angles of incidence that emits from the two LEDs. The angular-spectral information is then recovered from this intensity variation. Since the sensor pixels of size 6.45μm, there are 930 pixels in matrix **S** in a length of 6mm. In addition, the spectrum that effectively covers these two LEDs is chosen to be from 460nm to 640nm. With 5nm wavelength spacing, 39 wavelengths are considered. As a result, there are 31 × 39 = 1209 elements in **I**. The number of unknowns is larger than the number of measured values, making the inverse problem under-determined.

An iterative DBS algorithm serves to search for the solution with reasonable accuracy [16]. As mentioned earlier, matrix **I** contains the angular-spectral distribution of the object to be solved. Each element in **I** represents the intensity of light emitting from the object at a wavelength and an angle. In each iteration, elements in **I** are picked one by one in a random order and then a perturbation, either negative or positive, is applied to each element. The perturbation is kept or discarded based upon whether the metric defined below is minimized or not. The next iteration starts after all the elements in **I** are traversed. A random guess on **I** is considered as the initial solution. The entire algorithm terminates when either a maximum number of iterations is reached ($N_{max}$ = 200) or the metric converges to an acceptable value. The metric we used is the 2-norm of the residual error $\|\mathbf{S} - (\mathbf{AQ})\mathbf{I}\|_2$. The iterations terminated when the difference in this 2-norm between the current and the previous iterations was smaller than a threshold of $10^{-5}$. Figure 6 shows a typical curve of metric convergence. The algorithm is similar to our previous studies in computational spectrometer [16] and computational color camera [18], except the fact that unknown matrix **I** contains angular-spectral distribution instead of only spectral information. Also, singular-value-decomposition-based algorithm is able to efficiently reconstruct the unknowns without iteration [17,18], which will be explored in the future.

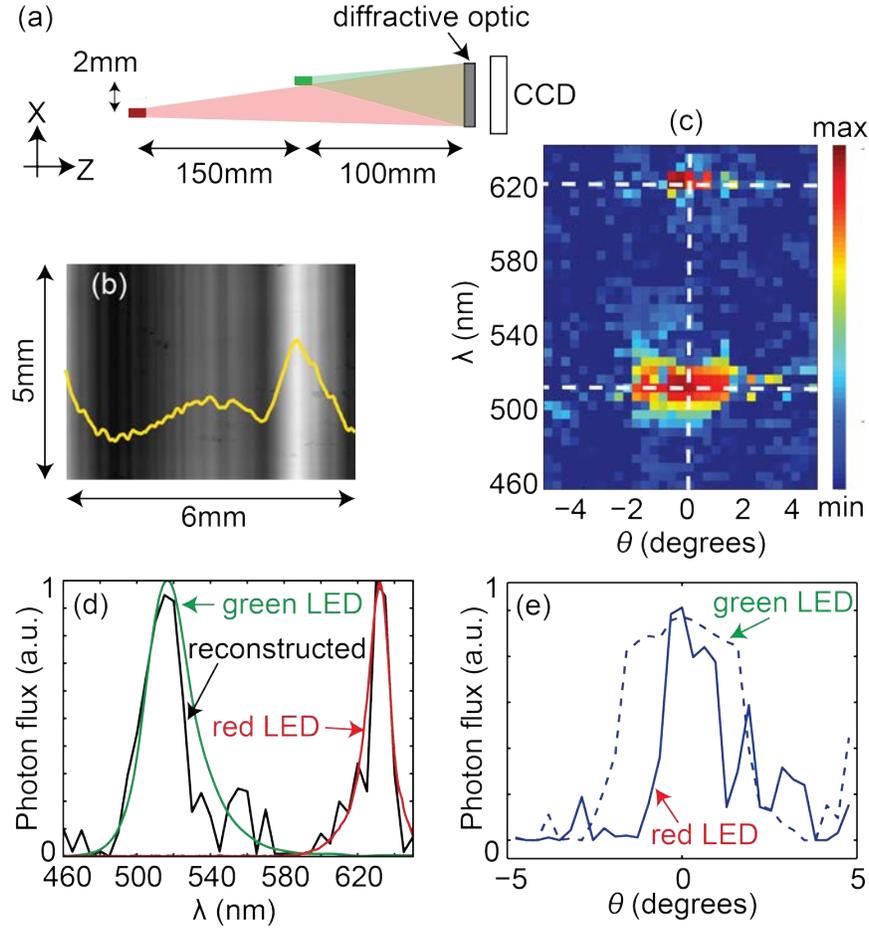

Fig. 5. (a) Schematic of the imaging experiment setup. (b) Raw CCD image with overlaid 1D intensity plot. (c) Reconstructed angular-spectral intensity distribution. (d) Reconstructed spectrum (black solid line) along vertical dashed lines in (c) at normal incidence. Green and red solid lines are reference spectra of two LEDs measured by a commercial spectrometer. (e) Reconstructed angle plot along top (red LED) and bottom (green LED) horizontal dashed lines at spectrum peaks in (c).

Figure 5(c) gives the reconstructed angular-spectral distribution, **I** from -4.8° to 4.8° and from 460nm to 640nm. Additionally, Fig. 5(d) plots the reconstructed spectrum along the dashed vertical line at normal incidence in Fig. 5(c). Evidently, there are two peaks at 515nm and 625nm, respectively, corresponding to the emission peaks of the two LEDs. The profiles of the reconstructed spectra match the reference spectra (green and red solid lines) measured by a commercial spectrometer with reasonable accuracy. The fluctuations are ascribed to errors in calibration, raw image and numerical reconstruction. Note that the LEDs used in this experiment have poor quality, for example spatial non-uniformity and low output power. This also contributes to the errors in the reconstruction. The recovered angular distributions for both green and red LEDs (along the top and bottom dashed horizontal lines in Fig. 5(c)) are plotted in Fig. 5(e). Both LEDs exhibit limited ranges of angles, which are basically determined by the width of the diffractive optic and the distance between the point sources and the diffractive optic. The red and green LEDs cover the angles from -0.32° to +0.96° and from -1.6° to +1.6°, respectively. This range is determined by applying a threshold of half of maximum to the plots in Fig. 5(e). The minimum and maximum angles are related to the coordinates of point source by $\tan(\theta_{min}) = ((L/2)-x)/z$ and $\tan(\theta_{min}) = ((L/2)+x)/z$. Here $L = 6$mm is the width of one diffractive optic in the X direction. Finally, we are able to make use of the angular distributions to retrieve the original point source. The X coordinate is $x=(L/2)\times(\tan(\theta_{min})+\tan(\theta_{max}))/(\tan(\theta_{min})-\tan(\theta_{max}))$, and the Z coordinate in longitudinal direction is $z=L/(\tan(\theta_{min})-\tan(\theta_{max}))$. The position of green LED is (0, -107mm) and the position of red LED is (1.5mm, -268mm). The refocusing accuracy is within 16% error. This can be significantly improved by using finer angular calibration and suppress the aforementioned factors that contribute to reconstruction errors. We acknowledge that since we are not collecting the complete hyper-spectral lightfield, there is ambiguity in depth estimation. In our experiment here, we are taking advantage of the fact that we have apriori knowledge that the scene is only comprised of 2 point-like sources to estimate the depth (locations of the point sources).

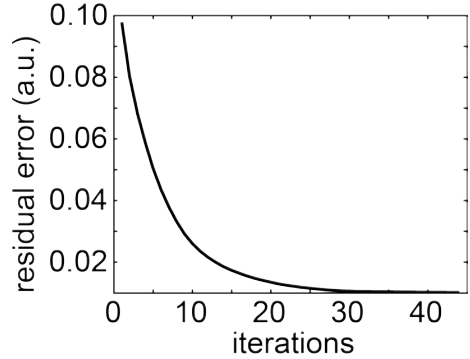

Fig. 6. Residual error convergence plot.

5.2 Spectral and angular correlation analysis

Finally, we analyzed the angular and spectral resolutions of the imaging system using the correlation function [16,23,24]. In order to estimate spectral resolution, we calculate the correlation between the AS-PSFs at two wavelengths spaced by $\delta\lambda$ at a fixed angle and a fixed spatial location of fiber-tip (see Fig. 3). It is expressed by [18]:

$$C_\lambda(x', \delta\lambda, \theta) = \frac{\langle PSF(x',\lambda,\theta) \cdot PSF(x',\lambda+\delta\lambda,\theta) \rangle_\lambda}{\langle PSF(x',\lambda,\theta) \rangle_\lambda \cdot \langle PSF(x',\lambda+\delta\lambda,\theta) \rangle_\lambda}. \quad (1)$$

Here, PSF($x'$, $\lambda$, $\theta$) represents the experimentally calibrated intensity at incident angle, $\theta$, wavelength, $\lambda$ and along X' direction on the sensor. It is also denoted as angular-spectral PSF. Additionally, $\langle...\rangle_\lambda$ means averaged over $\lambda$. A number of different $\delta\lambda$ values were computed. The correlation function $C_\lambda(x', \delta\lambda, \theta)$ is normalized. Multiple correlation functions at different incident angles and fiber-tip positions are averaged and the final data $C_\lambda^{avg}(\delta\lambda)$ is plotted in Fig. 7(a). As before, correlation decreases as the wavelength spacing $\delta\lambda$ increases. Using a correlation value of 0.5 as a threshold, we estimated spectral resolution of ~6nm.

Note that the distance $d$ between the diffractive optics and the sensor is the primary factor that dictates the spectral resolution for a given sensor. The shorter this distance, the lower the spectral resolution. Here, $d$ = 90mm was chosen so as to sufficiently distinguish the two LEDs. A much shorter distance will make it inaccurate to locate peaks of LED spectra and tell them apart. And a longer distance makes the system more bulky without adding much enhancement in spectral resolution.

A similar calculation is carried out for angular correlation, plotted in Fig. 7(b), which predicted an angular resolution of ~0.15°. Likewise, the angular correlation function is expressed by [18]:

$$C_\theta(x', \lambda, \delta\theta) = \frac{\langle PSF(x',\lambda,\theta) \cdot PSF(x',\lambda,\theta+\delta\theta) \rangle_\theta}{\langle PSF(x',\lambda,\theta) \rangle_\theta \cdot \langle PSF(x',\lambda,\theta+\delta\theta) \rangle_\theta}. \quad (2)$$

$\langle...\rangle_\theta$ means average over $\theta$ and Fig. 6(b) is the average of multiple correlation functions at different wavelengths and fiber-tip positions. This leads to about 5% to 15% error in positioning LEDs in both X and Z directions by taking the derivative of the expressions for both X and Z coordinates. This is consistent with the measured 16% error. Again, the plot is normalized. Note that another set of calibration data **A** is measured with finer angle (0.05°) and wavelength (1nm) steps to calculate the correlation functions shown in Fig. 7. But fewer spatial locations of fiber-tip are scanned and then averaged to save time. Optimization of both the diffractive optic and the system geometry will improve both the spectral and angular resolutions.

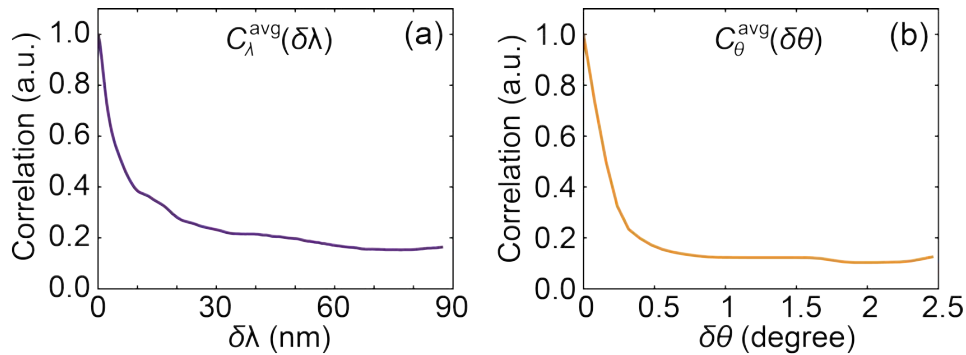

Fig. 7. (a) Spectral correlation versus wavelength spacing. (b) Angular correlation versus angle spacing.

5.3 Discussion

The form factor of the proposed imaging system is primarily limited by the distance between the diffractive optic and the sensor ($d$=90mm here). This distance can be significantly reduced by using diffractive optic of smaller feature size (< 1μm) [18]. Additionally, the size can be further minimized by optimizing the height profile [16,20,21] so that imaging resolutions are not compromised at a smaller distance. Our system is also lensless. Most cameras require complex lens assemblies. In addition, the conventional angle-sensitive camera requires a microlens array [4]. Coded aperture spectral imager needs to incorporate relay optics and a dispersion element such as a prism [13]. Nevertheless, we point out that our system is not able to recover occlusions.

The range of wavelengths that can be reconstructed are limited by three factors: (1) sensitivity of the semiconductor sensor; (2) bandwidth of the light source in calibration and (3) bandwidth of the spectrometer in calibration. The current system is able to cover the entire visible to near-IR spectrum (400nm – 1100nm). The incident angles that can be reconstructed are constrained by: (1) angular responsivity of the sensor and (2) finite acceptance angle of the fiber in calibration. The current system is able to cover a range of roughly ±20°. The experimental demonstration in this work falls far below these limits.

## 6. Simulation of 2D imaging

To demonstrate 2D angular-spectral imaging, a 2D diffractive optic is simulated [25]. For simplicity, we used a random height profile with square pixels of width 1μm and maximum height 1μm. We used a periodic structure comprised of 100 by 100 pixels. Its height profile is plotted in Fig, 8(a) and a magnified view of the right top corner is given in Fig. 8(b). Due to the smaller pixel size, the distance between the image sensor and the diffractive optic is only $d$=0.5mm.

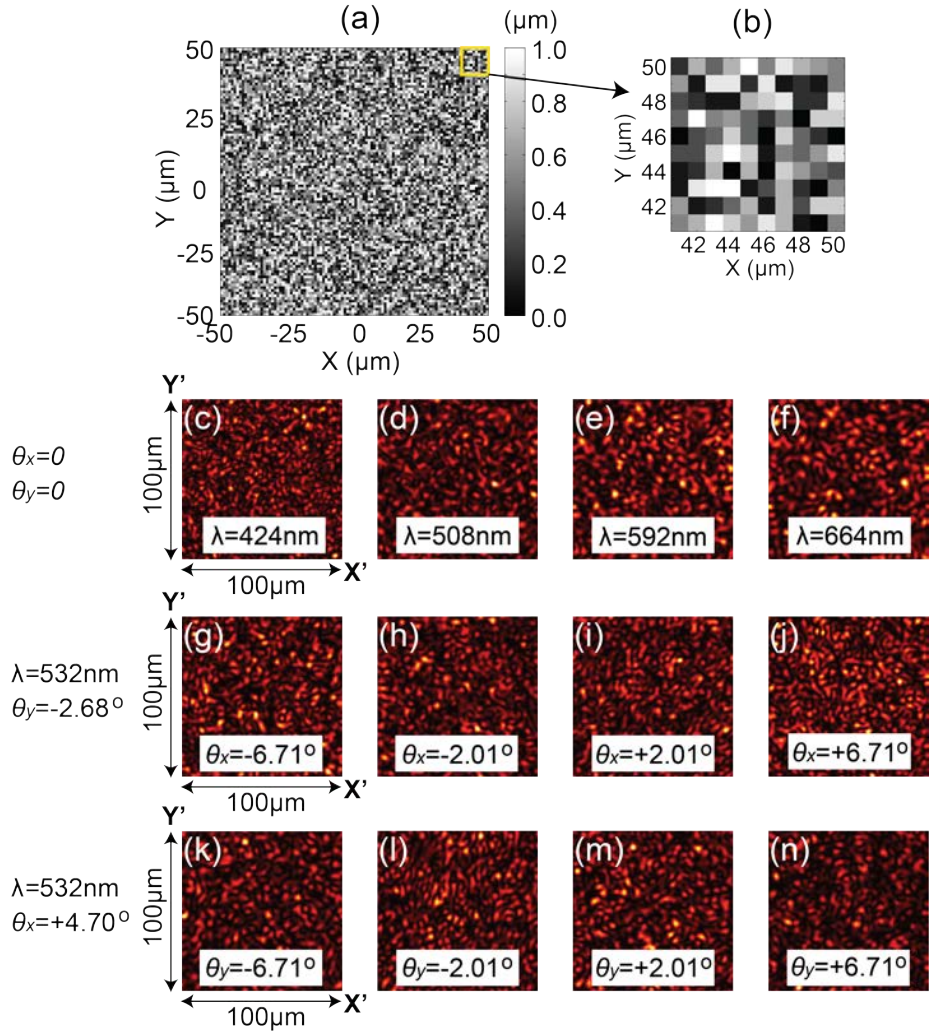

Fig. 8. (a) Height profile of the 2D diffractive optic. (b) Magnified view of the height profile of the right top corner of the 2D diffraction optic, enclosed in yellow square in (a). (c) – (f) Simulated diffraction patterns at four wavelengths when the angles of incidence are $\theta_x$=0 and $\theta_y$=0. (g) – (j) Simulated diffraction patterns at four angles $\theta_x$ when the wavelength is 532nm and $\theta_y$=-2.68°. (k) – (n) Simulated diffraction patterns at four $\theta_y$ when the wavelength is 580nm and $\theta_x$=+4.70°.

To characterize the system, the diffraction patterns of the diffractive optic were first simulated [18,20]. Note that different from previous studies where only normal incidence was considered [18,20,21], over-sampling of the device is necessary for oblique incidence. The rule-of-thumb for over-sampling rate is:

$$\Delta = \frac{\lambda_{min}}{4 \cdot \sin(\alpha_{max})}. \tag{3}$$

In Eq. (3), $\lambda_{min}$ is the minimum wavelength of interest and $\alpha_{max}$ is the maximum angle of incidence. Here, the spectrum of interest is from 400nm to 700nm with 12nm spacing, leading to 26 wavelengths. The ranges of angles in both X and Y directions are from -8.05° to +8.05°, with 0.67° spacing. Thus, there are 25×25 angles. Figures 8(c)-(f) are the simulated diffraction patterns at $\lambda$=424nm, 508nm, 592nm and 664nm, respectively, when $\theta_x$=0 and $\theta_y$=0. Similarly, Figs. 8(g)-(j) show different diffraction patterns at different angles $\theta_x$ at fixed $\lambda$=532nm and $\theta_y$=-2.68°. And Figs. 8(k)-(n) are diffraction patterns at different angles $\theta_y$ at fixed $\lambda$=580nm and $\theta_x$=4.70°. As anticipated, the angular-spectral PSFs change over wavelengths and angles of incidence. A spectral resolution of 9.35nm, and an angular resolutions of 0.11° in both X and Y directions can be estimated.

We then simulated a simple scene, which consists of 4 fluorescent beads (Alexa Fluor series) at 4 distinct locations in 3D. The field of view is 100μm×100μm. The imaging volume is located 1mm away from the diffractive optic and it is 1.4mm in depth. The spatial distributions of the beads are depicted in Fig. 9(a) and their emission spectra are plotted in Fig. 9(b). Usually the beads are <300nm in diameter, which can be treated as point sources. Each bead is assumed to emit spherical waves, impinging on the diffractive optic. Diffraction patterns of different wavelengths and angular frequencies are multiplexed and recorded by the sensor, which also has pixel size of 1μm. DBS is again applied for reconstruction. Note that there are 100×100=10000 measurements (knowns) and 25×25×26=16250 variables (unknowns). The reconstructed angular-spectral data are summarized in Fig. 9(c). Different beads occupy different locations and ranges of incident angles. Beads closer to the diffractive optic (B1 & B4) have broader angular intensity distributions, while those further away (B2 & B3) have narrower ones. As shown in Fig. 9(b), the residual error decreases and eventually converges as the algorithm iteratively progresses. A closer look at the reconstructed data, compared against the ground truth, is shown in Figs. 9(e) – (h). These are cross-sections of the 3D data cube. As can be seen, the plots of spectral (when both angles $\theta_x$ and $\theta_y$ are fixed) and angular (when both wavelength $\lambda$ and one of the angles $\theta_x$ and $\theta_y$ are fixed) intensity distributions match well with the true values.

Based on the reconstructed angular-spectral data, we can recover the 3D spatial coordinates of each bead. The X and Y coordinates are related to the minimum and maximum angles in both directions by $x=(L/2)\times(\tan(\theta_{xmin})+\tan(\theta_{xmax}))/(\tan(\theta_{xmin})-\tan(\theta_{xmax}))$ and $y=(L/2)\times(\tan(\theta_{ymin})+\tan(\theta_{ymax}))/(\tan(\theta_{ymin})-\tan(\theta_{ymax}))$. The Z coordinate is computed by $z=L/(\tan(\theta_{xmin})-\tan(\theta_{xmax}))$. The true coordinates of the beads are (0, 0, -1300), (-45, -45, -1800), (-25, 35, -2400) and (30, 15, -1000). And the coordinates from reconstruction are (0, 0, -1067), (-50, -50, -1564), (-30, 16.7, -2277) and (27.8, 13.7, -861), respectively. The coordinates are in micrometers. The percentage position accuracy in X, Y and Z directions are 9.6%, 18.1% and 12.5%, respectively. The inaccuracy is primarily ascribed to finite angular resolution and numerical errors in reconstruction. Using smaller angle calibration step size and improved reconstruction algorithms, such as by applying regularization [17], it is possible to improve performance.

By applying correlation functions to the simulated diffraction patterns, both spectral and angular resolutions can be obtained. Figure 10 illustrates the impact of $d$ on resolutions. Five values are included, $d$=25μm, 50μm, 100μm, 200μm and 500μm. Both spectral and angular resolutions drop as $d$ decreases and this happens in a highly nonlinear manner. This offers us confidence in the potential of reducing the size of the system without compromising resolution. Here, the resolutions are good for most common applications even when $d$ is as small as 0.1mm.

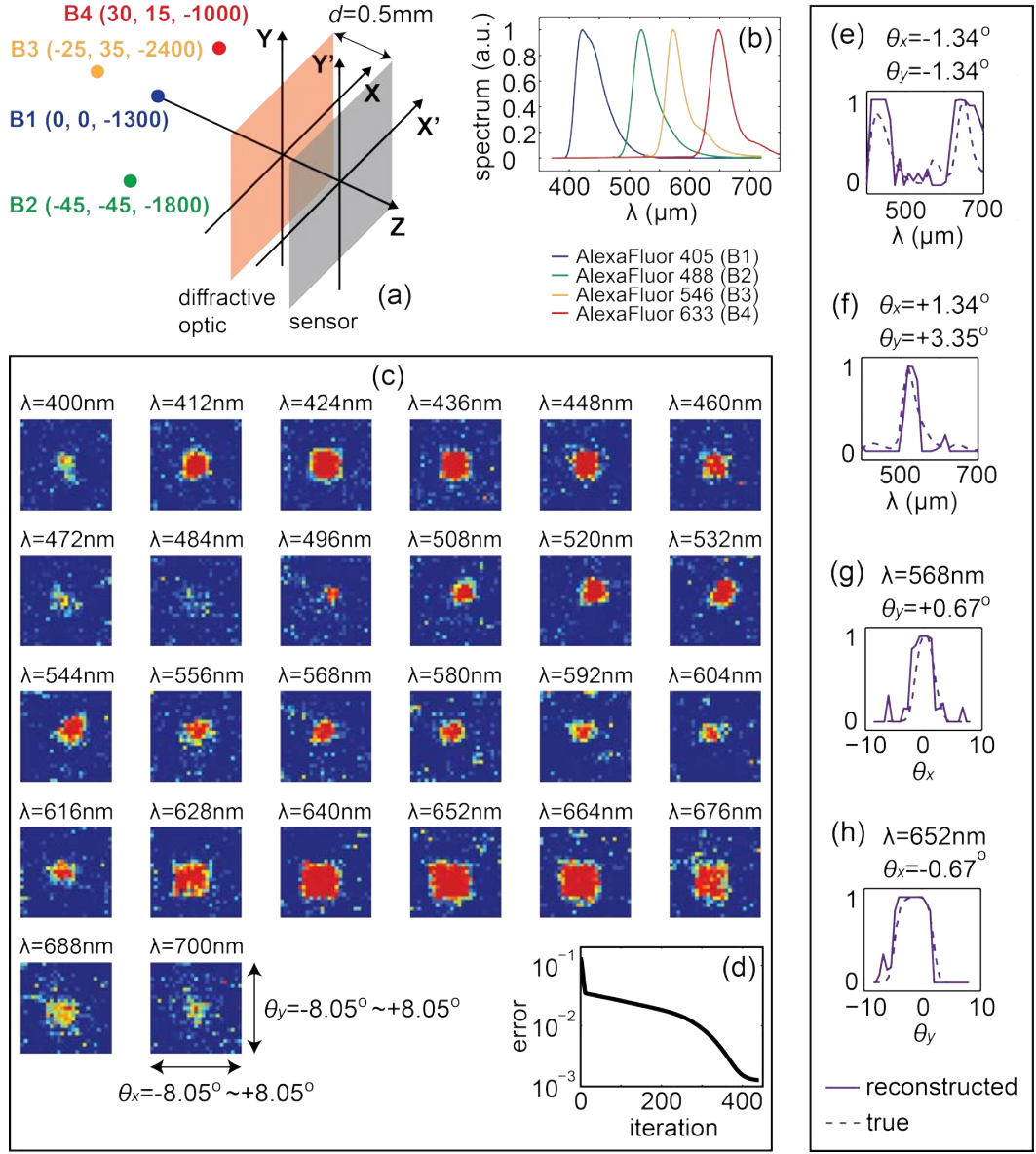

Fig. 9. (a) Schematic of the simulated imaging system comprised of the diffractive optic and sensor. Distributions and spatial coordinates of the four beads are labeled (in micrometers). (b) Emission spectra of the 4 fluorescence beads. (c) Reconstruction results using DBS. They are angular intensity distributions ($\theta_x$ and $\theta_y$ in horizontal and vertical directions, respectively) at 26 different wavelengths. (d) Evolution of residual error versus iteration number. (e) – (h) Plots of example data from reconstruction (solid purple lines) and ground truth (dashed purple lines) for (e) $\theta_x$=-1.34° and $\theta_y$=-1.34°, (f) $\theta_x$=+1.34° and $\theta_y$=+3.35°, (g) $\lambda$=568nm and $\theta_y$=+0.67°, and (h) $\lambda$=652nm and $\theta_x$=+0.67°.

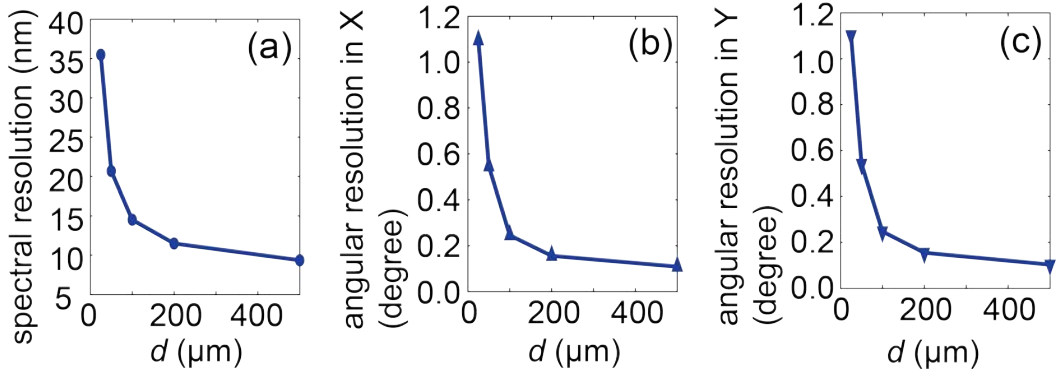

Fig. 10. Simulated (a) spectral resolution, (b) angular resolution in X, and (c) angular resolution in Y as a function of $d$.

## 7. Conclusions

In conclusion, this paper experimentally demonstrated a snapshot angular-spectral lensless camera. It is enabled by both a diffractive optic fabricated by single-step lithography and an iterative algorithm to solve the inverse problem. No lens is required, since the diffractive optic is employed to simultaneously distinguish both angular and spectral properties of the light. Thus it can be smaller than traditional imaging systems. It is capable of retrieving the angular-spectral distribution of a simple object (two point sources of two colors) with reasonable accuracy. Additionally, photon utilization rate is maximized due to the highly transparent diffractive optic. Since the computation is fast, video-rate imaging should be feasible as well. Here, the frame rate is only limited by the speed of the sensor, since no scanning is needed. Recently, there have been many efforts to encode multi-dimensional information onto single 2D image and then numerically recover the unknown object information [10,13,26,27]. Our single-element lensfree, planar configuration could potentially lead to a compact, low-cost, multi-dimensional camera. It is noted that sub-wavelength nanophotonic structures [28] can be incorporated into the diffractive optic to image the vector properties of light, such as polarization.

**Acknowledgments**

The project was funded by a DOE Sunshot Grant, EE0005959; a NASA Early Stage Innovations Grant, NNX14AB13G; and Office of Naval Research, 55900526.

The University of Utah has applied for patent protection for the subject technology. RM is the co-founder of Lumos Imaging, which has a commercial interest in the subject technology.